\documentclass[article,twocolumn]{aastex631}
\pdfoutput=1 %for arXiv submission
\usepackage{amsmath,amstext}
\usepackage{apjfonts}
\usepackage{xcolor}

\def\et3{\eta_3}
\def\th1{\theta_{-1}}
\def\r07{r_{0,7}}
\def\x05{x_{0.5}}

\def\TeV{\hbox{~TeV}}

\usepackage{xspace}

\def\thegrb{GRB~180720B\xspace}
\def\sword{{\it SWORD}\xspace}
\def\Fermi{{\it Fermi}\xspace}

\begin{document}

\title{Polarization Measurement of Gamma-ray Bursts with {\it Fermi}-GBM: The Case of GRB 180720B}
\author[0000-0002-2149-9846]{P.~Veres}
\affiliation{Department of Space Science, University of Alabama in Huntsville, Huntsville, AL 35899, USA}
\affiliation{Center for Space Plasma and Aeronomic Research, University of
Alabama in Huntsville, Huntsville, AL 35899, USA}
\author{W.~Duvall}
\affiliation{Space Science Division, U.S. Naval Research Laboratory, Washington, DC 20375, USA}
\author[0000-0002-0587-7042]{A.~Goldstein}
\affiliation{Science and Technology Institute, Universities Space Research Association, Huntsville, AL 35805, USA}
\author[0000-0003-2105-7711]{M.~S.~Briggs}
\affiliation{Center for Space Plasma and Aeronomic Research, University of
Alabama in Huntsville, Huntsville, AL 35899, USA}
\affiliation{Department of Space Science, University of Alabama in Huntsville, Huntsville, AL 35899, USA}
\author[0000-0002-0586-193X]{J.~E.~Grove}
\affiliation{Space Science Division, U.S. Naval Research Laboratory, Washington, DC 20375, USA}

\begin{abstract}
To achieve confident non-zero polarization measurements for gamma-ray bursts (GRBs) we need sensitive polarimeters and bright GRBs.
Here we report on the polarimetric analysis of the bright GRB 180720B using the \Fermi Gamma-ray Burst Monitor (GBM). We rely on the detection of photons that scattered off Earth's atmosphere and into GBM from this burst. 
Polarized gamma-rays will exhibit a characteristic pattern when scattering off the atmosphere that differs from an unpolarized beam.
We compare the measured photon counts in the GBM detectors with extensive simulations of polarized beams to derive the most probable polarization degree (PD) and angle (PA). 
For the entire GRB, we find PD$=72^{+24}_{-30}\% ~(1\sigma)$ and PA$=91^{+11}_{-9}$ deg ($1\sigma$, equatorial frame). Interestingly, the PA value is broadly consistent with an early optical PA measurement by the Kanata telescope, starting shortly after the end of the prompt emission. The consistency of PAs lends support for this method. The relatively high polarization degree (albeit with large uncertainties) agrees with similar past measurements suggesting that some GRBs might be highly polarized. This will be confirmed or refuted by the upcoming dedicated GRB polarimeters.
\end{abstract}

\section{Introduction} \label{sec:intro} 
Gamma-ray bursts (GRBs) are one the most luminous transients emitting the most of their energy in sub-MeV gamma-rays. GRBs have non-thermal spectra \citep{Band+93,Poolakkil+21spec} and one of the most natural non-thermal radiation mechanism is the synchrotron process. Synchrotron radiation, in turn, can lead to high intrinsic levels of polarization \citep{rybicki79} (PD$\lesssim$75\%).  Relativistic and geometric effects will alter the intrinsic polarization level, but still lead to significant (PD$\gtrsim$ 20\%) polarization for the observer \citep[e.g.][]{Toma+09pol}.

The level of polarization of the GRB prompt emission is an open question. There is no unambiguous, high significance detection of non-zero polarization at this time. AstroSat \citep{Chatto+09Astrosat}, INTEGRAL \citep{kalemci06,Gotz+14pol}, and IKAROS \citep{Yonetoku+12pol} measurements range from $PD\sim10\%$ to $100\%$, with significant uncertainties \citep[see e.g.][for a tabulation of the recent results]{Gill+20linpol}. On the other hand, measurements by POLAR \citep{Zhang+19Polar}  indicate a population of GRBs with low polarization levels. In general, the lack of significant non-zero polarization measurements are due to either low prompt polarization degree of GRBs or a lack of sufficient sensitivity of the current detectors. 
At this time there are a handful of dedicated gamma-ray polarimetry missions in different stages of planning (e.g. COSI \citep{Tomsick+22COSI} and POLAR-2 \citep{Hulsman20Polar2}). These will likely provide definitive measurements of GRB polarization. Until these missions materialize, we revisit the topic relying on currently available tools. Using the atmosphere as a scattering screen and relying on the polarization sensitivity of  Compton scattering, gamma-ray instruments with multiple detectors can be converted into polarimeters under fortuitous circumstances.

\begin{figure}
    \centering
    \includegraphics[width=\columnwidth]{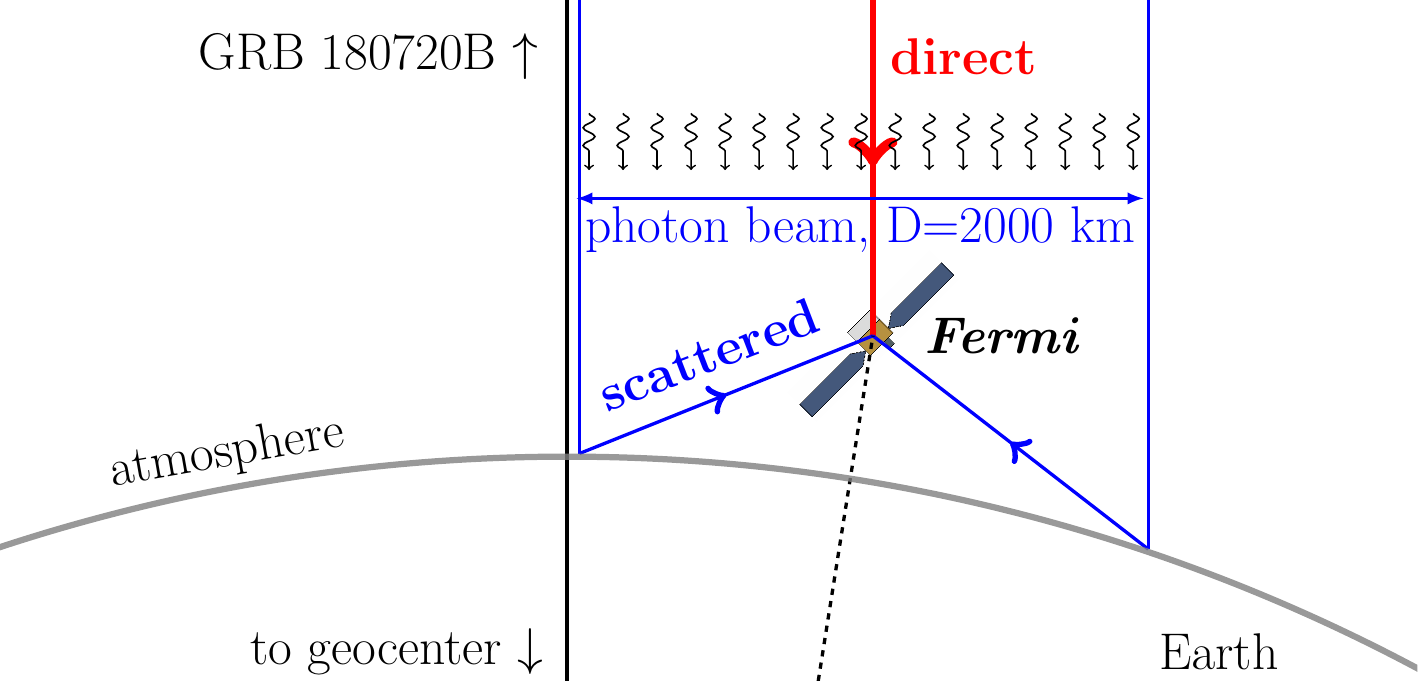}
    \caption{The scattering geometry showing the atmosphere (curved line), the photon beam of the simulations and the location of \Fermi.  The \Fermi -- geocenter -- GRB angle is 8.7 degrees. The altitude and position of \Fermi is to scale relative to Earth.}
    \label{fig:geom}
\end{figure}

The \Fermi Gamma-ray Burst Monitor \citep[GBM,\ ][]{meegan07} is the most prolific detector of GRBs \citep{vonKienlin+20GBM10yrcat}. Having 12 detectors pointed to cover a wide field of view, GBM monitors the entire unocculted sky for hard X-ray and gamma-ray transients like GRBs.
In observing geometries where the GRB occurs close to the local zenith of the spacecraft (Figure \ref{fig:geom}), 
Earth-facing detectors will have a non-negligible fraction of the flux from photons scattered off the atmosphere \citep{Pendleton+99locburst}.   
{GBM was designed to detect GRBs and characterize their time history and spectra. GBM detectors are not sensitive to polarization \citep{meegan07}.}

Photons preferentially scatter perpendicular to their electric field vector, imprinting a typical pattern on the location of the scattered photons (see e.g. Figure \ref{fig:scat_ex_atmo}, \citet{McConnell+94atmopol}).  We simulate this pattern by producing photon beams with a range of polarization degrees (PDs) and polarization angles (PAs).  Detectors pointing in different directions will be sensitive to this scattering pattern. 
The relative count level in the detectors will change as a function of PA and PD. We can match the counts from the simulated polarized photons to the observed counts and find the best fitting (PD, PA) solution. This method, and polarimetry in general, requires a large number of source photons (bright sources). \thegrb, which we analyze here, was one of the brightest GRBs detected by GBM and its geometry was favorable (zenith angle $\sim 9$ degrees) to look for polarization signatures.

This paper is structured as follows: in Section \ref{sec:obs} we describe the observations and the geometry we used. Next (Section \ref{sec:simulations}) we present the simulations of polarized photons and their comparison with the Fermi observations.  We present our results in Section \ref{sec:results} and conclude in Section \ref{sec:conclusion}.

\section{\thegrb observations}
\label{sec:obs}
\Fermi-GBM consists of 12 NaI detectors ({\tt n0}, {\tt n1}, \ldots {\tt nb}) sensitive in the 8-1000 keV range and 2 BGO detectors sensitive in the 0.2-40 MeV range.   The NaI detectors  point in different directions in order to cover the entire  unocculted sky.  Due to the observing schedule of \Fermi, it is common that  some detectors are pointed at or close to the Earth's atmosphere. Indeed this is how Terrestrial Gamma-ray Flashes \citep{Roberts+21MGF} are routinely detected. For gamma-ray detectors in low-Earth orbit, a fraction of the detected photons do not enter directly into the detector, but only after they scatter off the atmosphere.  For certain detectors, the scattered flux can even dominate over the direct flux. In the case of BATSE, for example, typically the scattered flux dominated over the direct flux starting from the detector with the third most counts \citep{Pendleton+99locburst}.

\thegrb triggered {\it Fermi}-GBM on August 20$^{\rm th}$, 2018 at 14:21:39.7 UT \citep[$T_0$,~][]{2018GCN.22981....1R}. It was also detected by {\it Fermi}-LAT \citep{2018GCN.22980....1B} and the Neil Gehrels {\it Swift} Observatory \citep{2018GCN.22973....1S,2018GCN.22975....1S}. The redshift of \thegrb is z = 0.654 \citep{vreeswijk+18180720bredshift}.  \thegrb was especially bright, and the afterglow was detected at very high energies (E$\gtrsim 0.1 \TeV$) by the H.E.S.S. telescope at late times  \citep{Abdalla+19HESS180720b}. 

The gamma-ray spectrum between T$_0$-1 s to T$_0$+56 s is best fit by the Band function \citep{Band+93}, which is a smoothly joined, broken power law (with indices $\alpha=-1.10\pm0.01$ and $\beta=-2.24\pm 0.03$). The peak energy in the energy-per-decade ($\nu F_\nu$) representation is $E_{\rm peak}=747\pm25 $~keV. The 1-second peak photon flux in the 10--1000 keV range is $124.5\pm0.6$ ph cm$^{-2}$ s$^{-1}$ and the  fluence (time integrated flux in the same energy range) is   $(2.9853 \pm0.0009) \times10^{-4}$ erg cm$^{-2}$  \citep{Poolakkil+21spec}. This is the 14$^{\rm th}$ highest fluence out of approximately 3800  GRBs detected by GBM in 16 years of observations.

We analyze the GBM data using the  {\tt RMfit}\footnote{\url{https://fermi.gsfc.nasa.gov/ssc/data/analysis/rmfit}} software.
We subtract a polynomial background fitted to intervals before and after the burst.  The time and energy integrated counts for each detector are displayed in Figure \ref{fig:sine} with dashed lines. We consider the 30--100 keV range when comparing the simulated and the observed flux because the scattered flux has the highest contribution in this range \citep{Willis+05baps}.

\begin{figure}
    \centering
    \includegraphics[width=1.0\columnwidth]{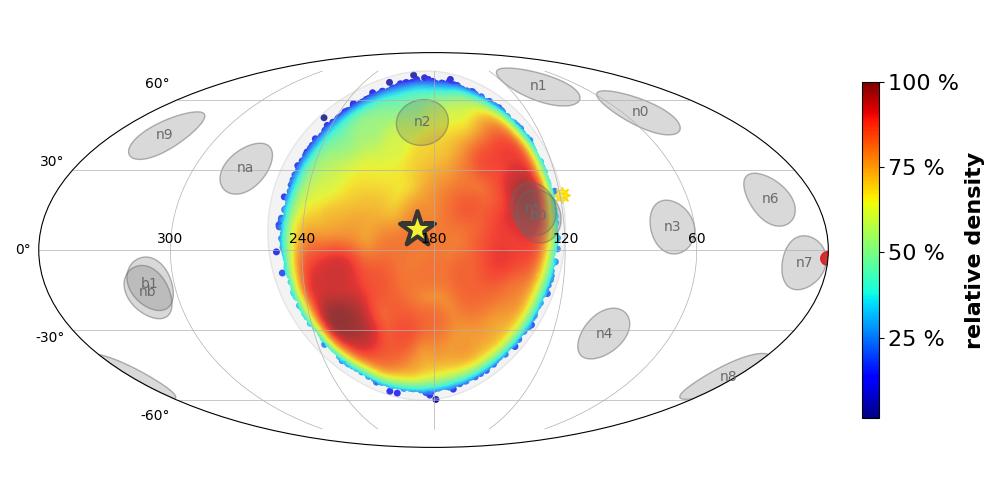}
    \caption{ Skymap at the time of \thegrb for \Fermi, Earth is the large circle at the middle of the figure with the geocenter indicated by the star. The intensity map shows the scattering locations on Earth of  100\% polarized (PA=0$^{\circ}$)
	    photon beam. Gray circles indicate the
	    pointing locations of the \Fermi-GBM detectors. The location of
	    \thegrb is indicated by a red circle.
    }
    \label{fig:scat_ex_atmo}
\end{figure}

{\it Geometry of the observation -} At the trigger time, the altitude of \Fermi was h=523.88 km (Earth radius R=6378.1 km), and the GRB -- geocenter -- \Fermi angle is 8.7$^{\circ}$ (Figure \ref{fig:geom}). Using the localization by {\it Swift}/XRT \citep{Page+18xrt180720b}
RA, dec = 00h02m06.86s, -02$^{\circ}$55'08.1'' (J2000) with uncertainty of 3.5'', we determined the angles of the detectors to the direction of the GRB and the geocenter (Table \ref{tab:det_pointings} and Figure \ref{fig:scat_ex_atmo}). Flux arriving directly from the GRB dominates in detectors $\tt{n3}$, $\tt{n6}$, $\tt{n7}$, $\tt{n8}$, $\tt{n9}$ and $\tt{nb}$.  The scattered flux from the atmosphere dominates detectors $\tt{n0}$, $\tt{n1}$, $\tt{n2}$, $\tt{n4}$, $\tt{n5}$ and $\tt{na}$. 

Detector $\tt{n7}$ is pointing closest to the GRB and we use this detector to normalize the {\it simulated, direct} flux to the observations. Detector $\tt{n4}$ has the least modulation among the detectors with significant scattered flux and we use this detector as the normalization of the {\it simulated, scattered} flux.

\begin{table}[]
    \centering
    \begin{tabular}{c|c c c}
Det. &  GRB & geocenter & pointing (RA, dec)\\ \hline
{\tt\bf n0} 	& 69.5  &      109.4 &  (48.1, 54.3)      \\ 
{\tt\bf n1} 	& 93.8  &      84.7 &   (93.3, 67.0)      \\ 
{\tt{\bf n2}} 	& 132.4  &      42.0&   (186.7, 49.9)     \\ 
{\tt n3} 	& 71.3  &      116.3&   (70.3, 8.6)       \\ 
{\tt{\bf n4}} 	& 92.3  &      98.0&    (94.1, -31.7)     \\ 
{\tt{\bf n5}} 	& 132.5  &      54.5&   (133.2, 15.4)     \\ 
{\tt n6} 	& 30.1  &      150.4&   (21.1, 18.8)      \\ 
{\tt n7} 	& 11.1  &      176.1&   (10.6, -4.7)      \\ 
{\tt n8} 	& 46.9  &      138.4&   (3.9, -49.5)      \\ 
{\tt n9} 	& 53.6  &      117.6&   (330.7, 44.3)     \\ 
{\tt{\bf na}} 	& 87.7  &      82.0&   (274.0, 30.8)      \\ 
{\tt nb} & 47.4  &      125.6&  (313.4, -15.7)     \\ 
    \end{tabular}
    \caption{Geometry of the detectors at the time of trigger showing the angle
    to the GRB, geocenter and the location of their pointing. Detector names in bold have favorable scattering geometry.}
    \label{tab:det_pointings}
\end{table}

\section{Simulations}
\label{sec:simulations}

\begin{figure}
    \centering
    \includegraphics[width=\columnwidth]{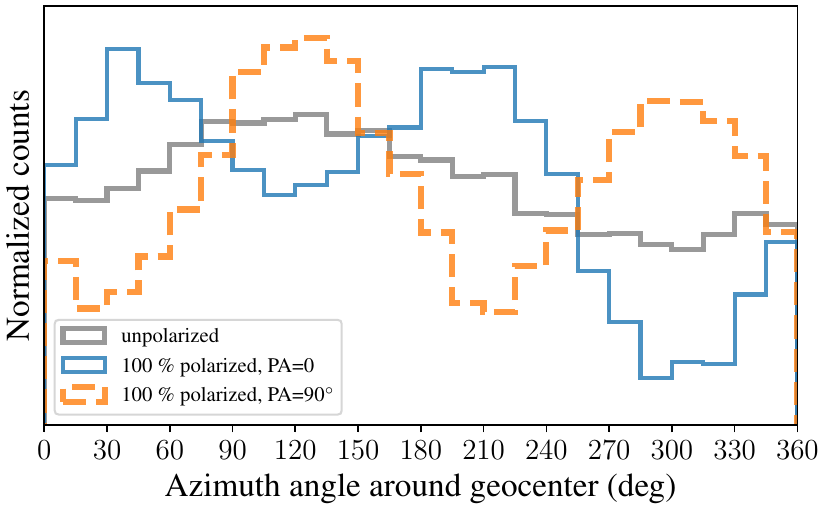}
    \caption{Modulation of the scattering pattern as a function of Azimuth
    angle around the geocenter for two cases of 100\% polarized with perpendicular polarization angles and an unpolarized beam. The modulation of the
    unpolarized beam results from the slight asymmetry of the GRB location.}
    \label{fig:scat_ex_modul}
\end{figure}

Using the atmosphere to investigate possible polarization signatures was proposed by \citet{McConnell+94atmopol} and applied to BATSE observations by \citet{Willis+05baps}. The method exploits the angular dependence of Compton scattering to linearly polarized photons. Photons preferentially scatter perpendicular to their electric field vector. If the gamma-rays are polarized, this will reflect in a pattern when plotting the last scattering location of the photons that enter the  \Fermi volume. Indeed simulations of linearly polarized photons show a strong modulation as a function of azimuth angle around the geocenter compared to the unpolarized beam (Figure \ref{fig:scat_ex_modul}).

We simulate photon beams using a grid of (PA, PD) values. We compare the appropriately normalized simulated counts in the Earth-facing detectors with the observed counts. We search the range of PA-PD values that are consistent across the selected detectors. We use $\chi^2$ statistics to perform the search, and the $\Delta \chi^2$ to determine the confidence region and we note that the $\chi^2$ is equivalent to the Z-statistic, used by \citet{Willis+05baps}.


\subsection{The \sword simulation software and NRLMSIS} 
The SoftWare for Optimization of Radiation Detectors (\sword) \citep{9032903} is a radiation transport  software developed at the U.S. Naval Research Laboratory for the Department of Homeland  Security. \sword provides a CAD-like interface to construct models and a large library of pre-existing  models, interfaces with Geant4 \citep{Agostinelli+03Geant4}, {MCNP \citep{Hendricks+11MCNP}} and {Omnibus \citep{Johnson11omnibus}} radiation transport engines, and also provides analysis  tools.

\sword handles polarized radiation through the Livermore Low Energy Electromagnetic Physics library in   Geant4 4.10.01.p02 \citep{Agostinelli+03Geant4}. 
To estimate the effect of atmospheric scattering on GRB observations, we built a realistic atmosphere based on the NRLMSISE-00 empirical model \citep{Picone+02nrlmsise} and simulate the transport  and scattering of bursts of gamma-rays with Geant4. NRLMSIS describes the  density and composition of the atmosphere from the ground through the thermosphere. While the deviation from a simple exponential density profile and the varying molecular composition with altitude described in  NRLMSIS have at best second-order effects on scattering, incorporating these details is straightforward and provides an easily referenced atmospheric standard.

The simulations consist of a photon beam of 2000 km diameter, centered around \Fermi from the direction of the GRB. \Fermi's volume is represented as a spherical detector of 80 km radius.   Photons  are injected (see Figure \ref{fig:geom}) with energy ranging from 10 keV to 10 MeV with a spectral  shape corresponding to the measured spectrum of \thegrb (Section \ref{sec:obs}). To construct the model of the atmosphere NRLMSIS was run with a resolution of 0.1 km from 0 to 1000 km. A \sword model was created using concentric  spherical shells. Each shell has a different density, pressure, and elemental makeup, using the average  from NRLMSIS covered by each shell. Shells have linear or logarithmic spacing and the shell spacing was picked to match a previous effort to simulate the atmospheric response of \Fermi \citep{fermi_sim}.

Because it is computationally expensive, we used beams that cover scattering up to 55$^{\circ}$ from the atmosphere, where the maximum possible angle is 67.5$^{\circ}$ (the angular extent of Earth as viewed from \Fermi). To have the correct comparison with the observed photons without affecting the polarization pattern, we extrapolated the simulated distribution out to 67.5$^{\circ}$ in azimuthal slices around the center of the atmosphere. This extrapolation preserves the azimuthally symmetric polarization pattern, and allows the consistent normalization between the observed and simulated flux (both direct and scattered components).

In the simulation, photons are detected both directly entering the \Fermi detector volume and after experiencing one or more scatterings in the atmosphere. We record the arrival time of both the direct and the scattered photons in the detector volume. The scattered photons are delayed compared to the direct photons by more than 2 ms which allows us to clearly separate the two populations.

We convolve the direct photons with the known detector response matrix of GBM corresponding to the direction of \thegrb. This yields the {\it simulated}, {\it direct} counts in each of the GBM detectors. {We note that the response matrices account for the scattering on the spacecraft structures and it is unlikely that such scattering induces a polarization signature.} For calculating the {\it simulated}, {\it scattered} counts, we tile the Earth's atmosphere into small  regions using the {\tt healpix} \citep{Gorski+05healpix,Zonca2019} scheme. We use NSIDE=8 parameter in {\tt healpix} that corresponds to a resolution of about 7 degrees. We investigated higher resolutions in a limited number of cases with consistent outcomes. We determine the direction of each  {\tt healpix} region and construct a response matrix for that direction. We convolve the scattered photons coming from the given {\tt healpix} region and calculate the detected counts by GBM. Next we sum all the contributions from different directions spanning the atmosphere to obtain the {\it simulated, scattered} counts. This calculation mimics the detection process of GBM, starting from physical photon spectrum and resulting in a detected count distribution.

Based on the physics of the Compton scattering, the scattered photons for a typical GRB spectrum will have energies predominantly in the 30--100 keV range. While the simulations are carried out over the entire GBM sensitivity range (10 keV--10 MeV) we restrict the comparison with observations to the 30--100 keV range. For this reason, the BGO detectors were not considered for atmospheric scattering, and we only used them to derive the shape of the GRB spectrum.

We simulate an unpolarized beam and 100\% polarized beams with polarization angles ranging over 0--180$^\circ$ in increments of 15 degrees. The partially polarized beams contain a mix from the unpolarized and a 100\% polarized beam with the appropriate polarization angle. The partial polarizations range from 0--100\% in steps of 10\%.

We start with a beam of $\sim10^9$ photons. For our geometry, we detect approximately 100,000 scattered photons for each polarization angle. Because of fluctuations we take 90,000 photons in each case for consistency.

\begin{figure*}
    \centering
    \includegraphics{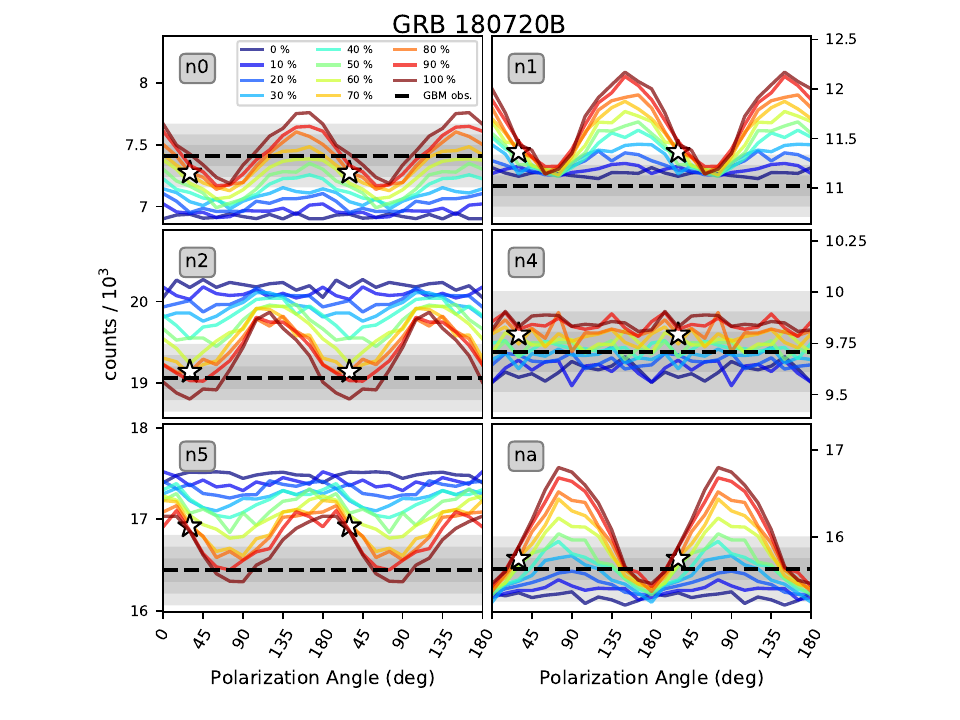}
    \caption{Modulation of the simulated photons in GBM. The colors
    indicate different polarization degrees. The dashed line shows the detected counts
    with the shaded region indicating the 1, 2 and 3$\sigma$ uncertainties
    (statistical only). Stars indicate the best $\chi^2$ value computed for all detectors by scanning the PA-PD grid (Figure \ref{fig:pdpa}), which is $({\rm PA}, {\rm  PD}) = (30^{\circ},\, 80\%)$. We extended the x-axis to show two periods for clarity. }
    \label{fig:sine}
\end{figure*}

We use the $\chi^2$ statistic to compare the simulations with the observations.
\begin{equation}
\label{eq:chi2}
\chi^2=\sum_{{det.}}\frac{(C_{\rm obs}-C_{\rm sim})^2}{\sigma^2_{C_{\rm obs}} + \sigma^2_{C_{\rm sim}}},
\end{equation}
where $C_{\rm obs}$ is the observed counts and $C_{\rm sim}$  is the simulated counts. $C_{\rm sim}$ is a function of the PA and PD, and $\sigma$ indicates the uncertainty. The summation is over the detectors with good scattering geometry, except detector {\tt n4}, which is used for normalization.

\subsection{Normalization to observations}
\label{sec:simnorm}
To apply the method of \citet{McConnell+94atmopol,Willis+05baps}, a key step is finding the scaling factors that relate the simulated direct and scattered counts to the observed counts in \Fermi. 
We normalize the {\it direct} counts of the simulations using detector {\tt n7}, which points closest to the GRB. The scattered flux is normalized using the detector that has significant scattered flux and the least amount of modulation due to the changing polarization angle (see detector {\tt n4} in Figure \ref{fig:sine}). 

The observed number of photons in each detector is comprised  of scattered and direct components.
After finding the direct and scattered normalization factors  using detectors {\tt n7} and {\tt n4} respectively, we find (as a sanity check) that the normalized, simulated counts are  within 5\% of the observed values ({\tt n4} in Figure \ref{fig:sine}).

\begin{figure}
    \centering
    \includegraphics[width=\columnwidth]{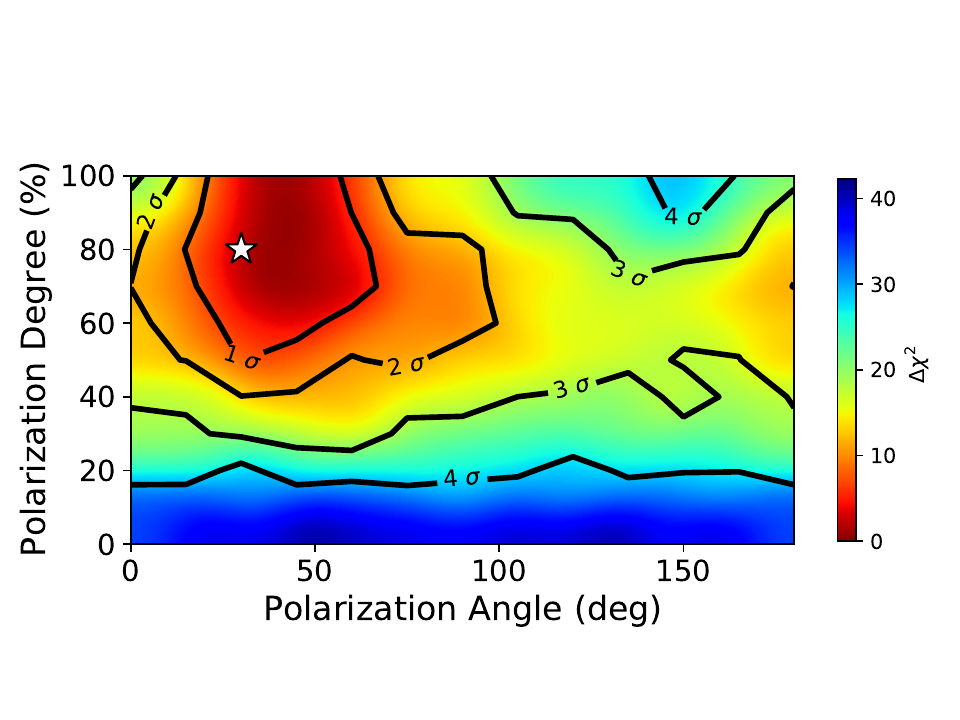}
    \caption{Polarization angle -- polarization degree diagram for \thegrb, based on comparison of simulations to the observed data in 5 detectors. Star marks the best solution, $({\rm PA}, {\rm  PD}) = (30^{\circ}, 80\%)$, consistently on each of the panels.   } 
    \label{fig:pdpa}
\end{figure}

\subsection{Sources of systematic error}
\label{sec:sys}
The uncertainty on both the simulations and the observed counts suffer from systematic terms in addition to the Poisson statistical uncertainty. Below we list the most important sources of systematic uncertainty.

%\begin{itemize}
1.) The spectral response of GBM does not include a model of the solar panels. Solar panels affect a small portion of the field of view in 5 out of 6 detectors with good scattering geometry. One of the \Fermi solar panels however is in the sensitive region of detector {\tt na} as it views the atmosphere. To assess the possible effects of the solar panel on detector {\tt na}, we calculate the polarization estimate without detector {\tt na} and find consistent results, meaning the solar panel blockage is not significant.  {Based on the study of the solar panel position we estimate a systematic error of at most 5\% in the number of simulated counts due to the solar panels. }

2.) When selecting the 30-100 keV range, GBM's fixed 128 channel energy edges  will not have exactly the desired energy edges for this range. We select the channels that include the  30 or the 100 keV limits, which introduces a minor systematic uncertainty. {By including (or excluding) an additional energy channel at the lower limit (30 keV), we estimate at most 1 \%  uncertainty in the observed counts.}

3.) The shape of the GRB spectrum is uncertain, resulting from e.g. spectral evolution, or just simply by the deviations of the real spectrum from the fitted spectral function. {We estimate about 5\% uncertainty based on fitting spectral shapes (e.g. power law with an exponential cutoff or a smoothly broken power law) that represent a worse approximation than the Band function reported in Section \ref{sec:obs}.} 

4.) When we extrapolate in azimuthal sections out to the edge of the atmosphere, we also introduce some systematic uncertainty. {Based on the uncertainty of the fit parameters used in the extrapolation, we estimate a systematic error of $\lesssim$10\%.}

5.) Because the actual atmosphere is not static, its properties will be slightly different from the model we assumed in the \sword simulations. {It is difficult to estimate the changes in density in different atmospheric layers compared to our model at the time of \thegrb, and their effect on scattering gamma-rays. Density models of different atmospheric layers can be uncertain by $\approx10\%$ \citep[e.g.][]{Yu+22atmo}. We therefore estimate the contribution of this component to be at most $10\%$.}

6.) The simulations covered energies up to 10 MeV. The flux above this threshold from \thegrb is negligible, but still may contribute to scattered flux. {By extrapolating the spectrum above 10 MeV and estimating the relative flux from this range into the 30-100 keV interval, we estimate the systematic uncertainty $\lesssim 1\%$  }

7.) We subtract the background based on pre- and post burst intervals, fitting a polynomial. Even though the background surrounding \thegrb is well fit by a polynomial, the interpolation of the background during the time of the GRB will introduce a minor uncertainty. {We estimate this uncertainty based on the background fit polynomial parameters to be $\approx1\%$. }

{8.) The detector responses were measured and validated \citep{2007AIPC..921..590K,2008AIPC.1000..565H} using a detailed model of the spacecraft. The complex interior structure of {\it Fermi} however might involve scattering effects that are not fully modeled and introduce systematic uncertainties. While it is unlikely that the scattering on the spacecraft elements introduces a polarization signature, the mis-identification of photon energy might result in a systematic uncertainty.  Considering the ratio of effective areas of direct and scattered flux between 30 and 100 keV for the selected Earth-facing detectors (Figure \ref{fig:eff}), we estimate an uncertainty of at most $5\%$. }

For our purposes, we will assume the contribution of systematic errors is $50\%$ of the statistical error budget and quote the errors based on this assumption. We stress that such a large systematic does not flow from GBM instrumental uncertainties or the  \sword simulations, but rather from  the novelty of the method, and the desire to be conservative.

\section{Results}
\label{sec:results}
We perform a comparison between the observed counts in the detectors with significant scattered flux and the simulated counts based on the (PA, PD) parameter grid. Using the $\chi^2$ statistic in Equation \ref{eq:chi2}  between the observed and simulated counts we find the best (PA, PD) solution.
We display the simulated and normalized counts for a range of PDs as a function of PA in the 6 detectors with significant scattered flux in Figure \ref{fig:sine}.

Figure \ref{fig:sine} also conveys the essence of the method: extracting polarization measurements using an instrument not intrinsically sensitive to polarimetry. The modulated signature typical of dedicated gamma-ray polarimeters \citep[e.g.][]{Gotz+09varpol} is only present in our simulated observations, because the simulations are the only component of our analysis where we can explicitly probe the polarization. Whereas standard polarimeters show modulation in, for example, the scattering angle distribution, here we have modulation as a function of the polarization angle. The panels in Figure \ref{fig:sine} show the polarization-sensitive simulations (colored lines) for all possible (PD, PA) combinations. We search for the (PD, PA) combination that best matches the observed count rates (dashed lines) across all detectors. In summary, by utilizing a grid of polarization-sensitive simulations, it is the relative photon counts combined with the observing geometry that makes GBM sensitive to polarization.

We find that the simulation that best matches the observations for  5 detectors (excluding {\tt n4}) has PD=80\% and  PA=30 deg ($\chi^2=6.8$ for 5 degrees of freedom). We note that the solution, marked by a star indicates consistent PA-PD over all detectors. The star is located on the PD=80\% line in all panels, and at the PA=30 deg. For detectors {\tt n1} and {\tt n5} the solution is at or above the 3$\sigma$ level (statistical only). We attribute this discrepancy to the presence of systematic uncertainties (see Section \ref{sec:sys}) and account for these in Figure \ref{fig:pdpa}.

In Figure \ref{fig:pdpa} we present the confidence region in the PA-PD plane based on $\Delta \chi^2$ relative to the best solution. The size of the 1$\sigma$ region depends on the assumption about the systematic errors. In Figure \ref{fig:pdpa} we have assumed the baseline case where the systematic error is half of the statistical error. The PA-PD map shows the PD=0\% case can be ruled out with $\gtrsim 4\sigma$ significance. We note that if we assume that the systematic error equals the statistical error, the PD=0 case can be ruled out at $\gtrsim 3\sigma$ significance level (the best solution does not change).
After marginalizing over the polarization angle, the polarization degree is $PD=72^{+24}_{-30}$\% ($1\sigma$). We similarly marginalize over the PD to measure the polarization angle and get $PA=35^{+11}_{-9}$ deg ($1\sigma$). Thus, we find that in order to explain the level of scattered flux in the GBM detectors, we need to assume a relatively high polarization degree.

\section{Discussion}
\label{sec:conclusion}

\subsection{Comparison to previous work}
\citet{Willis+05baps} applied the method for only 2 out of the 8 BATSE detectors. In this paper we used data from 5 out of the 12 GBM detectors to search for polarization signatures.  In our case, any 2 of the 5 detectors with good modulation signal could be used similarly to \citet{Willis+05baps}. As a sanity check, we searched the 10 pairs of detectors for polarization signal, and they all give solutions consistent with (PA,\ PD)=(30$^{\circ}$,\ 80\%), with the exception of  the {\tt n1}-{\tt na} pair. The reason for the discrepancy is that the phase difference for this pair is close to 90$^{\circ}$ (see Figure \ref{fig:sine}), essentially producing no change in the $\chi^2$ goodness of fit as we vary the PA and PD. {We note that while the {\tt n1}-{\tt na} pair cannot constrain the polarization, together with other detectors both {\tt n1} and {\tt na} provides meaningful constraints.} We {also} note that using only 2 detectors at a time results in significantly larger confidence regions. 

\subsection{Interpretation}

Because of its high luminosity, and no rising afterglow signature at long wavelengths (characteristic of off-axis geometries), \thegrb was likely observed on-axis. This means the viewing angle to the jet symmetry axis, $\theta_v$ is smaller than the opening angle of the jet, $\theta_j$. Among the possible emission models, synchrotron radiation emitted in a volume filled with ordered magnetic field is the best candidate to account for such a large polarization degree \citep{Toma+09pol}.   We numerically integrated equations 4 and 5 of \citet{Toma+09pol} to calculate the observed Stokes parameters (and the PD), in an ordered magnetic field scenario. We consider the observed photon index of \thegrb and assume we are viewing the jet approximately at half the opening angle ($\theta_v=0.5\theta_j$). We find that the maximum polarization obtainable is $PD\sim 35\%$, which is  consistent within 1.2$\sigma$ of the measured 72\%.

An alternative scenario that can result in higher polarization is where the jet is viewed close to its edge, within $1/\Gamma$ (where $\Gamma$ is the Lorentz factor) or $\theta_j - 1/\Gamma \lesssim \theta_v<\theta_j + 1/\Gamma$ \citep{Granot+03highpol021206}. Here we assume a top-hat jet, and for the spectral parameters of \thegrb we find the maximum $PD\sim 55\%$, for $(\Gamma \theta_j)^2\approx 2$. This scenario still results in $PD\gtrsim 40\%$ for $(\Gamma \theta_j)^2\lesssim 300$. Thus, a reasonable opening angle, $\theta_j\approx 0.1 $ and a Lorentz factor of 100 can accommodate this scenario and account for the large PD within $1\sigma$.

\subsection{Early optical polarimetry for \thegrb}
Interestingly, the Kanata optical telescope \citep{Sasada+18Kanata} followed up  \thegrb and measured early polarization  at the few \% level  (starting at T$_0$+70 s, whereas our observations end at T$_0$+56 s). The PA measured by Kanata in the T$_0$+70 s to 300 s interval   \citep{Arimoto+23_180720b} is $\sim70$ degrees (50-80 degrees range), measured in the  equatorial system. For comparison, our PA measurement ($PA\approx35$ degrees, \sword frame) corresponds to $PA=91^{+11}_{-9}$ degrees in the  equatorial frame (56 degree shift). The early optical polarization, if ascribed to the reverse shock, likely probes the same ejecta that produces the gamma-rays. Given the systematic uncertainties of the gamma-ray method and the fact that the two measurements are not coeval, we consider the closeness of the PAs a favorable indication that the method we presented is indeed measuring real polarization. 

\subsection{\it Future prospects}
Indirect polarimetric measurements such as the method presented here require a specific source geometry, namely a small zenith angle for the GRB. In addition, the GRB needs to have a large flux. Based on approximately 10 years of \Fermi data \citep{vonKienlin+20GBM10yrcat}, we estimate that the rate of GRBs with sufficiently high peak flux (P$>$10 ph cm$^{-2}$ s$^{-1}$) that occur within 20 degrees of the local zenith of \Fermi is $\sim$2\%. This means that approximately 2\% of \Fermi GRBs are suitable for this type of analysis, or in excess of 70 GRBs. If we relax the zenith angle cut to 30 degrees but require a higher flux of 30 ph cm$^{-2}$ s$^{-1}$, the number of suitable GRBs will be close to 50. Because the detectors might not point sufficiently close to the atmosphere for some of these GRBs, each has to be treated on a case-by-case basis.

In this paper we presented a proof-of-concept measurement of polarization using the atmosphere as a scattering environment, for an instrument that was not designed for polarimetry.  Even though the high degree of polarization claimed here comes with a relatively large error, the measurement is valuable as there are only a handful of polarimetric observations.  At this time it is unclear if the GRB prompt emission, in general, will show high polarization degree ($PD\gtrsim 50\%$) or if the polarization degree is at the level of few percent. With dedicated gamma-ray polarimetry missions such as POLAR-2 \citep{Hulsman20Polar2} and COSI \citep{Tomsick+22COSI} (at $\gtrsim$0.2 MeV energies), we are entering the era of routine polarimetric  GRB observations where this question will be answered.

{\bf Acknowledgements: } {We thank the anonymous referee for thoughtful questions.} We thank Makoto Arimoto for sharing their early optical polarization results on \thegrb before publication. We acknowledge discussions with Mark McConnell, P\'eter M\'esz\'aros {and Narayana Bhat}. The authors acknowledge Fermi GI grant 80NSSC20K0414.  P.~V. and M.~S.~B. gratefully acknowledge NASA funding from cooperative agreement 80MSFC22M0004.  A.~G. gratefully acknowledges NASA funding through cooperative agreement 80NSSC24M0035. 

{\it Software}: numpy \citep{vanderWalt+11numpy}, matplotlib \citep{Hunter07matplotlib}, astropy \citep{astropy+18}, scipy \citep{Scipy+20}, GBM data tools \citep{GbmDataTools}, healpix \citep{Gorski+05healpix}, healpy \citep{Zonca2019}.

\appendix

\begin{figure}
    \centering
    \includegraphics[width=\columnwidth]{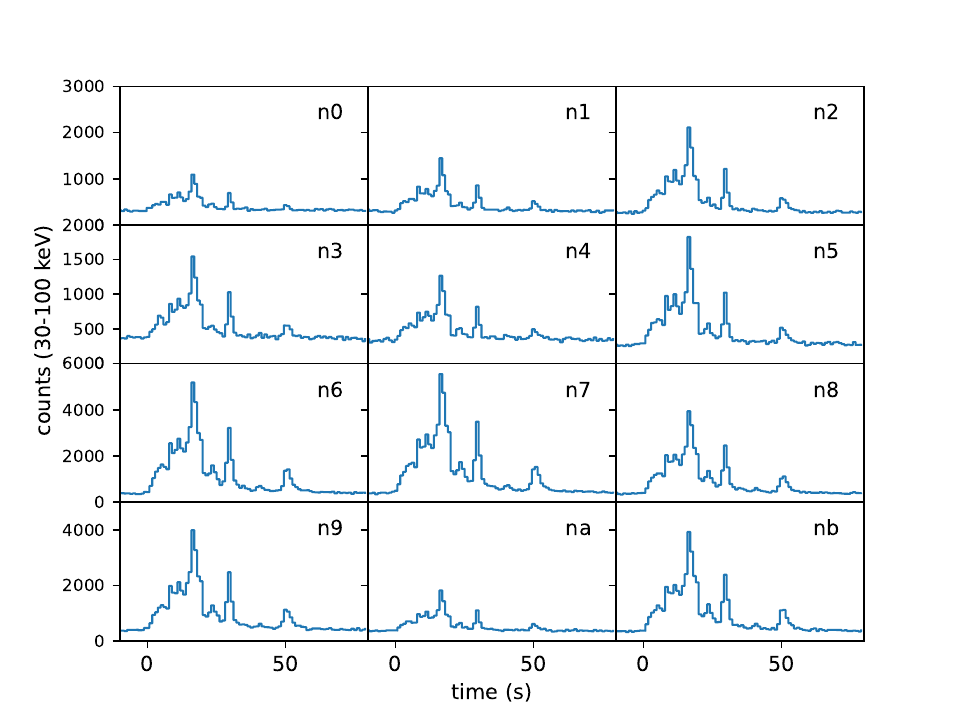}
    \caption{\bf Lightcurve of \thegrb for all NaI detectors in the 30-100 keV range. The temporal resolution is 1 s. Panels in every given row have the same y-range.}
    \label{fig:lc}
\end{figure}

\begin{figure}
    \centering
    \includegraphics[width=\columnwidth]{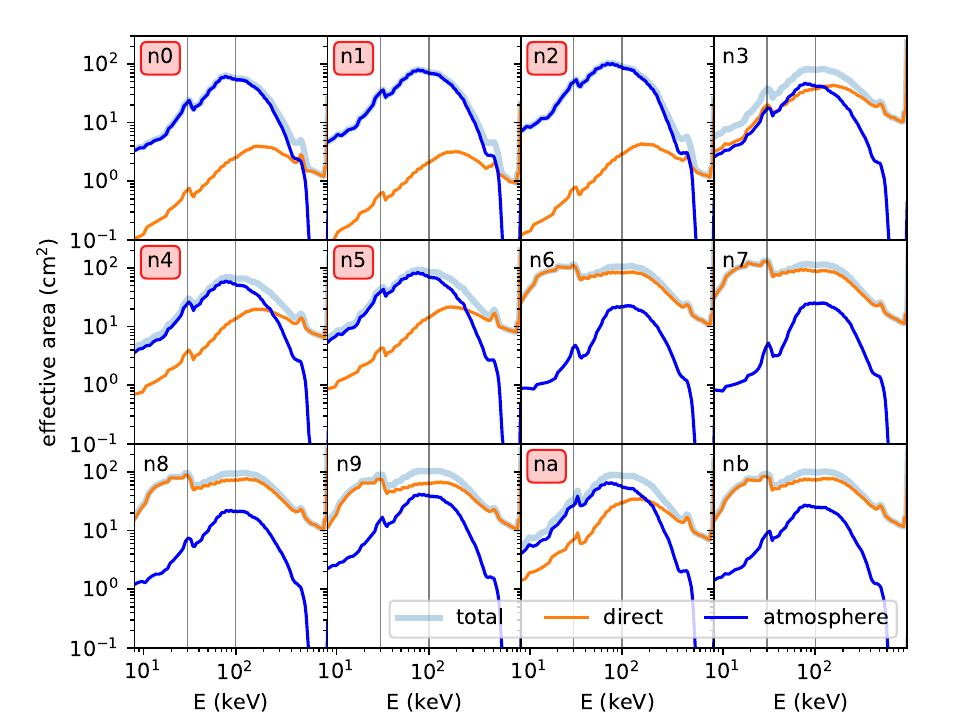}
    \caption{\bf Comparison of the detector responses for the NaI detectors showing the effective area. We indicate the direct and atmospheric contribution for each detector. The horizontal lines mark the interval used in this study (30-100 keV). Detectors marked in pink were used for the polarization analysis.}
    \label{fig:eff}
\end{figure}

\bibliographystyle{yahapj}
%\bibliography{GBMatmo,grb,polarization}

\end{document}